\documentclass[runningheads]{llncs}
\usepackage{graphicx}
\usepackage{subcaption}
\usepackage{amsmath}
\usepackage{uppaal}
\usepackage{xcolor}
\usepackage{xspace}
\usepackage[T1]{fontenc}
\usepackage{array}
\usepackage[utf8]{inputenc}
\usepackage{cite} 

\begin{document}
\title{Safe Failure Fraction Verification and Optimization in Industrial Drive Systems}
\titlerunning{Safety Verification and Optimization in Industrial Drive Systems}
%

\author{Imran Riaz Hasrat\inst{1} \and
Eun-Young Kang\inst{1} \and
Christian Uldal Graulund\inst{2}}


%
\authorrunning{Hasrat et al.}
%
\institute{The Mærsk Mc-Kinney Møller Institute, Software Engineering Department,\\
University of Southern Denmark (SDU), Odense, Denmark\\
\email{\{imrh,eyk\}@mmmi.sdu.dk}
\and
Danfoss Drives A/S, Ulsnaes 1, DK-6300 Gråsten, Denmark\\
\email{christian.graulund@danfoss.com}}
\maketitle              
\begin{abstract}

Safety and reliability are crucial in industrial drive systems, where hazardous failures can have severe consequences. Detecting and mitigating dangerous faults on time is challenging due to the stochastic and unpredictable nature of fault occurrences, which can lead to limited diagnostic efficiency and compromise safety.
This paper optimizes the safety and diagnostic performance of a real-world industrial Basic Drive Module~(BDM) using {\sc uppaal stratego}. We model the BDM’s functional safety architecture with timed automata and formally verify its key functional and safety requirements through model checking to eliminate unwanted behaviours.
Considering the formally verified correct model as a baseline, we leverage the tool’s reinforcement learning facility to optimize the safe failure fraction to the 90\,\% threshold, improving fault detection ability. The promising results highlight strong potential for broader safety applications in industrial automation.

\end{abstract}
\section{Introduction}

In the current fast-paced industrial world, industrial drive systems play a crucial role in automating processes in diverse industries like automotive, aerospace, and food and beverage~\cite{sectors}. In recent years, the demand for smooth automation and precision has steadily increased making these systems the backbone of industrial manufacturing for controlling speed, torque, and direction in the operations of machines.  Since the drives primarily control the operation of electric motors, safety and reliability have become the top considerations for flawless and correct operation. 
However, the complex architecture (involving power converters and diagnostic means) of such drives poses challenges in achieving these goals.

Functional safety standards such as IEC 61508 and EN 61800-5-2 require robust safety-related design techniques to be applied in industrial applications. Safe Torque Off~(STO) is an important concept in shielding motors against faults—deactivating the motors to avoid situations that can lead to accidents and ensuring safety. This not only demands solid hardware design but also sophisticated diagnostic mechanisms to detect and react to failures in real-time.

Formal verification~\cite{FV} is a powerful technique for ensuring the functional correctness of complex real-time industrial systems. The safe failure fraction (SFF) metric is used to evaluate these systems' ability to detect failures. However, the randomness of fault occurrences makes optimizing SFF challenging, emphasizing the need for advanced techniques. Advancements in artificial intelligence make reinforcement learning~(RL)~\cite{sutton2018reinforcement} a strong candidate for optimization problems, such as enhancing SFF. RL can improve fault detection mechanism and safety by enabling systems to learn and adapt.

This paper aims to optimize SFF performance in an industrial BDM drive through a real-world case study. To do so, we use an RL-enabled timed automata (TA) based tool {\sc Uppaal Stratego}~\cite{stratego}, widely known for formal modelling, verification (using model checking), and optimization. In the first step, we construct the model of the BDM architecture as a set of TAs. We selected TA as the formal modelling technique due to their ability to represent complex systems with discrete states and time-dependent behaviours, making them ideal for the BDM, where timing and synchronization are critical.
Next, we perform formal verification against the functional and safety properties of the BDM to ensure its formal model is free from undesired behaviours and operates as intended. Formal verification ensures that the optimization process gets a correct model. Given the growing popularity of RL in solving optimization problems, we leverage the RL (Q-learning) capabilities embedded in {\sc Uppaal Stratego} to optimize the SFF performance.
The RL based optimization capabilities in {\sc Uppaal Stratego} make it well-suited for training the diagnostics of the BDM and enhancing overall safety.
The contributions of this study are as follows:

\begin{itemize}
    \item Formal modelling of the BDM architecture in {\sc Uppaal Stratego}, capturing its functional behaviour, safety mechanisms, and timing constraints.
    \item Formally verifying the functional and safety properties of the BDM, including core functionality and timing requirements, and ensuring the absence of undesired behaviours.
    \item Optimizing the SFF to achieve the target safety levels using RL in {\sc Uppaal Stratego}.
\end{itemize}

\section{Preliminaries}\label{pre}
In this section, we provide an overview of reinforcement learning and {\sc Uppaal Stratego}, which serve as the essential background for this paper.

\subsection{{\sc Reinforcement Learning}}
RL trains a machine how to behave by letting it interact with the environment. It computes the state-action pair values and collects the scores of the actions taken by the machine at different states. 
Q-learning~\cite{watkins1992q} is an RL algorithm that uses the Bellman-optimal equation to calculate the state-action pair values (called Q-values) by estimating the expected future reward for taking a specific action in a given state. The Q-learning algorithm can be represented as follows:
\begin{align*}
    Q^*(s, a) = E\left[R(s, a) + \gamma \max_{a'} Q^*(s', a')\right]
\end{align*}
Here, $Q^*(s, a)$ is the expected cumulative reward for taking an action $a$ from a state $s$, while $E$ denotes the expectation function over possible future states. $R(s, a)$ is the immediate reward received from the environment when performing an action $a$ from state $s$, $\gamma \in [0, 1]$ is the discount factor that adjusts the weight/importance of future rewards, and $\max_{a'} Q^*(s', a')$ is the maximum expected future reward achievable from the next state $s'$ by taking action $a'$.
\subsection{{\sc Uppaal Stratego}}
{\sc Uppaal Stratego}~\cite{stratego} is an advanced extension of the {\sc Uppaal} tool suite~\cite{uppaal1}, integrating traditional {\sc Uppaal} capabilities with RL techniques to derive safe and near-optimal strategies for Stochastic Hybrid Games~(SHG)~\cite{10.1007/978-3-031-46002-9_3}. 
{\sc Uppaal Stratego} is useful to generate strategies that can be used to optimize various performance metrics, such as minimizing cost or maximizing safety. 

 To synthesize optimal control strategies within a specified context, {\sc Stratego} incorporates RL algorithms like Q-learning. When applying Q-learning, {\sc Stratego} calculates Q-values by running sample runs, and these computed values are then used to improve the strategy. This iterative procedure is continually revised until the performance converges.
In the case of continuous state spaces, {\sc Stratego} employs an online partition refinement method to discretize the state space~\cite{10.1007/978-3-030-31784-3_5}. This technique allows the tool to dynamically enhance its representation of the state space during exploration. Consequently, effective strategies can be developed without requiring a comprehensive state analysis.
For an in-depth understanding of the tool's rich formalism, refer to~\cite{uppaal1,stratego,uppaal5}.
\section{Case Study}\label{case}
The case study describes the functional safety~(FS) architecture of an industrial BDM provided by Danfoss~\cite{drives}.
The BDM is an electronic power converter combined with its associated control system, positioned between an electric motor and power supply. 
It converts the supplied DC power into a format suitable for the smooth operation of the motor according to its design specifications.

\begin{figure}
    \centering
    \includegraphics[width=0.9\linewidth]{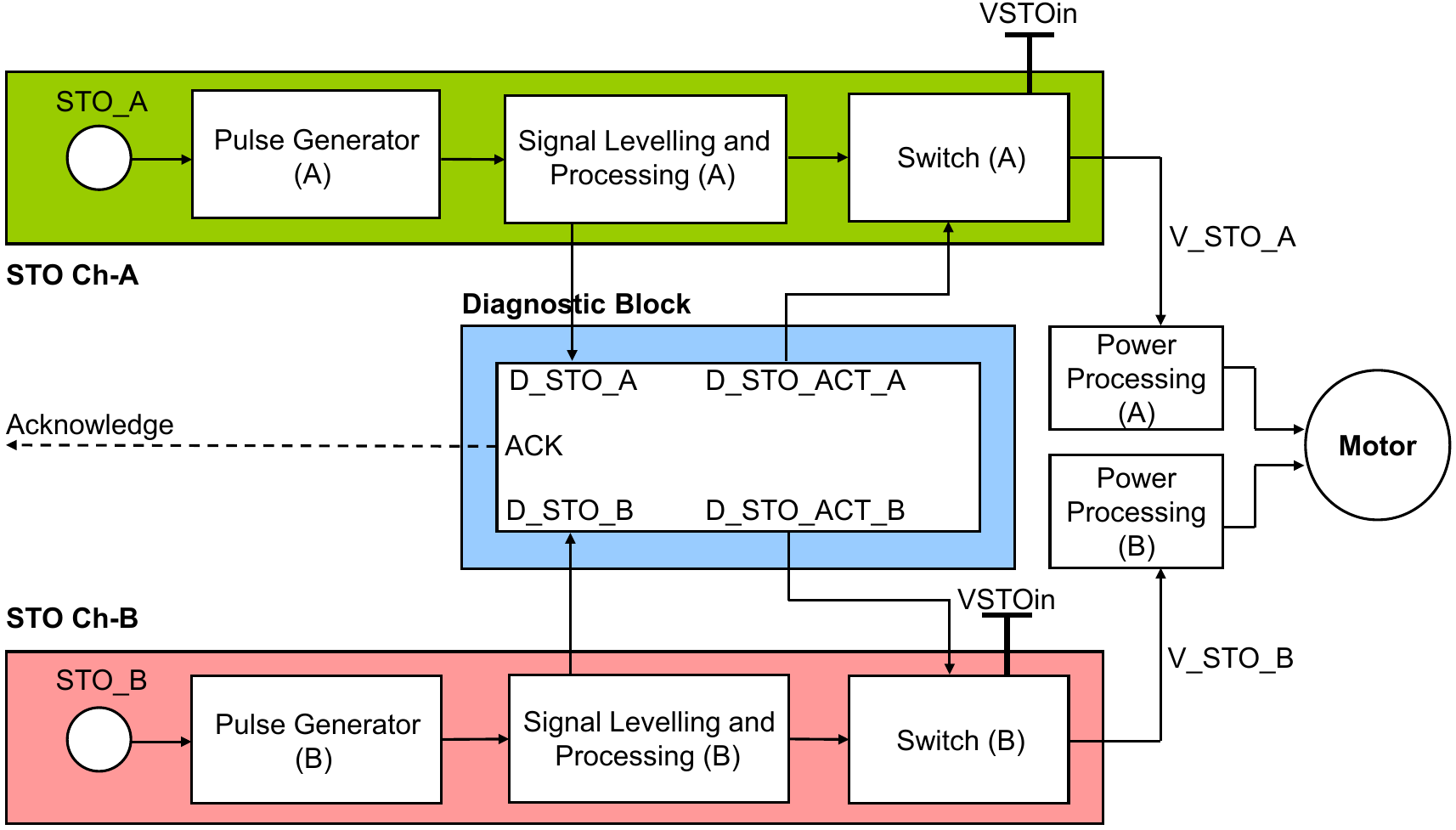}
    \caption{Overview of the functional safety architecture of the BDM}
    \label{fig:overview}
\end{figure}

The BDM is required to integrate the iSM3~(Safety Module), with an emphasis on implementing the SIL3~(Safety Integrity Level) STO function, along with internal diagnostics. To ensure an adequate safety level, the diagnostics in the iSM3 must achieve a SFF of at least 90\,\% as the threshold. The overview of the BDM is illustrated in Figure~\ref{fig:overview}. The functional blocks i.e., pulse generator, signal levelling and processing, switch, power processing and motor express the core functionality of the BDM.
On the other hand, the diagnostic block~(DB), being a non-functional block, implements diagnostic capabilities without directly contributing to the primary function of the BDM. We will now discuss the functionality of each block in detail.

\noindent {\textbf{STO Channels:}} The BDM comprises two safety channels, STO~Ch\_A and STO~Ch\_B, depicted in green and pink colours, respectively.
Each channel consists of three functional blocks identical to those in the other channel. The channels operate independently and provide redundancy and fault tolerance, supporting high safety integrity levels. Both channels are responsible for implementing the STO function, which prevents the unintended motion of the motor. 
The input power is supplied to the channels through the STO\_A and STO\_B terminals.
Whenever the STO function is activated, it ensures the de-energization of STO inputs which leads to removing the power supply to the motor, eliminating the risk of hazardous situations. 

\noindent {\textbf {Pulse Generator:}}
The pulse generator converts DC power supplied from the STO\_A and STO\_B terminals into a pulsating AC signal when the input voltage is within the normal operating range. 
Whenever STO is requested, the pulse generator gets de-energized from the STO\_A and STO\_B terminals.

\noindent {\textbf {Signal Levelling and Processing (SLP):}}
The implementation of STO in drives is publicly known, with a limited number of available practical methods detailed in an IFA report~\cite{IFA_available}. The general methods include main contactor cuts, pulse blocking, and others. While publicly known, Danfoss does not disclose its chosen method or the blocks involved. Instead, they are generalized as a single block, ``Signal Levelling and Processing.'' The pulsating AC signal passes through the SLP, enabling failure detection within its sub-blocks, before being converted back to DC. The DC output then controls the opening and closing of the switch based on the high/low value of the original signal.

\noindent {\textbf {Switch:}} 
The switches supply power to ``Power Processing'' blocks. STO follows the de-energize-to-trip principle, ensuring safety by opening the switches.

\noindent {\textbf {Power Processing:}}
The motor stays operational when "Power Processing" blocks receive continuous power. When the power supply is at logical low, the blocks turn the motor off, ensuring a safe state. Conversely, when high, the blocks turn the motor on, allowing normal function.

\noindent {\textbf {Diagnostic Block (DB):}}
DB continuously monitors channel blocks (through D\_STO\_A and D\_STO\_B) to diagnose faults. When a fault is detected, DB enforces a safe state by triggering channel switches (using D\_STO\_ACT\_A and D\_STO\_ACT\_B outputs) to cut power (V\_STO\_A and V\_STO\_B) to the ``Power Processing'' blocks. Additionally, DB sends an acknowledgement signal to the user interface, informing the operator that the motor is off and an STO request is active. The operator must then inspect, resolve the fault, and confirm the system is ready for operation again.

\section{Methodology}\label{meth}

Our methodological approach is divided into three phases: modelling, formal verification, and SFF optimization.

\noindent \textbf{1. Formal Modelling: }In this phase, we formally represent the structural and behavioural aspects of the BDM architecture in {\sc Uppaal Stratego}. The structural mapping is performed by defining semantics of each DBM block as a separate TA, ensuring a one-to-one correspondence between architectural components and their formal representations. Since the identical blocks in STO Ch\_A and Ch\_B perform the same tasks, they are grouped together and formally represented by a single TA with unique IDs.
On the other hand, behavioural mapping is performed by translating each block's dynamics, interactions, and synchronization mechanisms into formal semantics. This mapping approach is inspired by the work presented in \cite{10.1007/978-3-642-24270-0_18}. We now discuss the formalism and behaviour of single TA and the overall system.

A TA for any block is defined as a tuple $A=(L,\ell_0,C,D, \Sigma,T,I)$. Here, $L$ is the finite set of locations, and $\ell_0$ $\in$ L is the initial location. $C$ represents real-valued clocks, while $D$ defines the data variables. $\Sigma = \Sigma_c \cup \Sigma_u$  represents a finite set of actions, where $\Sigma_c$ are controllable and $\Sigma_u$ are uncontrollable. $T$ is the set of transitions/edges, where each transition is a tuple $(\ell,g,a,u,\ell')$, with $\ell$, $\ell'$ $\in$ L as source and destination locations, $g$ $\in$ \text{Guards}($C, D$) as a condition on clocks or data variables, $a$ $\in$ $\Sigma$ representing an internal action or synchronization with other blocks, and $u$ $\in$ \text{Updates}($C, D$) updating clocks or data variables. Finally, $I$ assigns invariants to locations, enforcing timing constraints on staying there.

The complete model comprises several TAs, called a network of TAs. In the BDM architecture, communication or information sharing between blocks is represented by arrows (see Figure~\ref{fig:overview}), while in the network of TAs, this happens through transitions using synchronization channels or shared variables. Formally, the network of TAs is defined as:
\begin{align*}
    N &= (A_1, A_2, \dots, A_n, V, C_h)
\end{align*}
\noindent where:
\begin{itemize}
    \item $A_i = (L_i, l_{0i}, C_i, D_i,\Sigma_i, T_i, I_i)$ represents the TA for block $i$.
     \item $V$ is the set of shared variables (and global clocks).
      \item $C_h$ is the set of synchronization channels.
\end{itemize}

\noindent \textit{Operational Semantics of Network of TAs:} The execution progress of the network at any given moment is defined by its global state $s$:
\begin{align*}
   s = (l_1, c_1, d_1, \dots, l_n, c_n, d_n)
\end{align*}
\noindent where $(l_i, c_i, d_i)$ represents the current location, clock valuations, and data variables of automaton $A_i$. The progress evolves through two types of transitions:

\noindent \textit{1. Time Delayed Transitions:}
Time progresses uniformly across all the TAs while maintaining location invariants:
\begin{align*}(l_1, c_1, d_1, \dots, l_n, c_n, d_n) \xrightarrow{t} (l_1, c_1 + t, d_1, \dots, l_n, c_n + t, d_n)\end{align*}
\noindent where t $\geq$ 0 and for each $i$, the condition $ci + t$ $\models$ $I_i(l_i)$ must hold.

\noindent \textit{2. Discrete Transitions:}
A TA $A_i$ transitions instantaneously when triggered by synchronization with any other block or when allowed by a guard condition:
\begin{align*}
(l_1, c_1, d_1, \dots, l_i, c_i, d_i, \dots, l_n, c_n, d_n) \xrightarrow{a} (l_1, c_1, d_1, \dots, l_i', c_i', d_i', \dots, l_n, c_n, d_n)
\end{align*}
\noindent if there exists a transition $(l_i, g, a, u, l_i')$ $\in$ $E_i$ such that:
\begin{itemize}
    \item $g(c_i, d_i)$ holds.
    \item $a$ is either a synchronization event or an internal action permitted by $I_i(l_i)$.  For synchronization to occur, another block $A_j$ must simultaneously execute a complementary transition: $a \in \Sigma_i \cap \Sigma_j$, ensuring coordinated execution.
    \item The update function $u$ modifies clocks and data to $c_i', d_i'$. 
\end{itemize}

\noindent \textbf{2. Formal Verification: }In the second phase, we utilize {\sc Stratego's} verification engine to formally verify the correctness of the constructed formal model of the BDM against a set of functional and safety requirements formulated as CTL~\cite{uppaal5} queries. For example, verifying motor deactivation upon fault detection and the correct sequence of actions in different blocks to ensure correct and safe operation under both normal and faulty conditions.

\noindent \textbf{3. SFF Optimization: }Once the formal verification step is complete, we optimize DB performance to achieve a target SFF of 90\,\% using the formally verified correct model as a baseline. First, we define an objective function to set the optimization criteria. 
Then controllable and uncontrollable actions are introduced to allow {\sc Stratego} understand what actions can be controlled. The network of TAs, with controllable and uncontrollable transitions, incorporating continuous clocks and stochastic fault occurrences, collectively forms an SHG. The RL facility in {\sc Stratego} needs to solve our SHG model to generate a winning strategy, which can be defined through the function  $\rho \to \Sigma_c' \subseteq \Sigma_c$. As the strategy approach presented in this paper is memoryless, we express it as $w: s \to \Sigma_c' \subseteq \Sigma_c$ where $s = \text{last}(\rho)$ is the last state of $\rho$. Conceptually, $w$  directs the “system” player to select controllable actions from the states that help to achieve the desired objective.  
\section{BDM Formal Modelling, Verification and SFF Optimization using {\sc uppaal stratego}}\label{mod}
In this section, we describe the formal modelling of BDM, the verification of its core functionality, and the optimization of SFF performance.
\subsection{Modelling the Behaviour of the BDM as Timed Automata}
\begin{figure*}[t!]
    \centering
    \begin{subfigure}{\textwidth}
        \centering
        \includegraphics[width=1.0\linewidth]{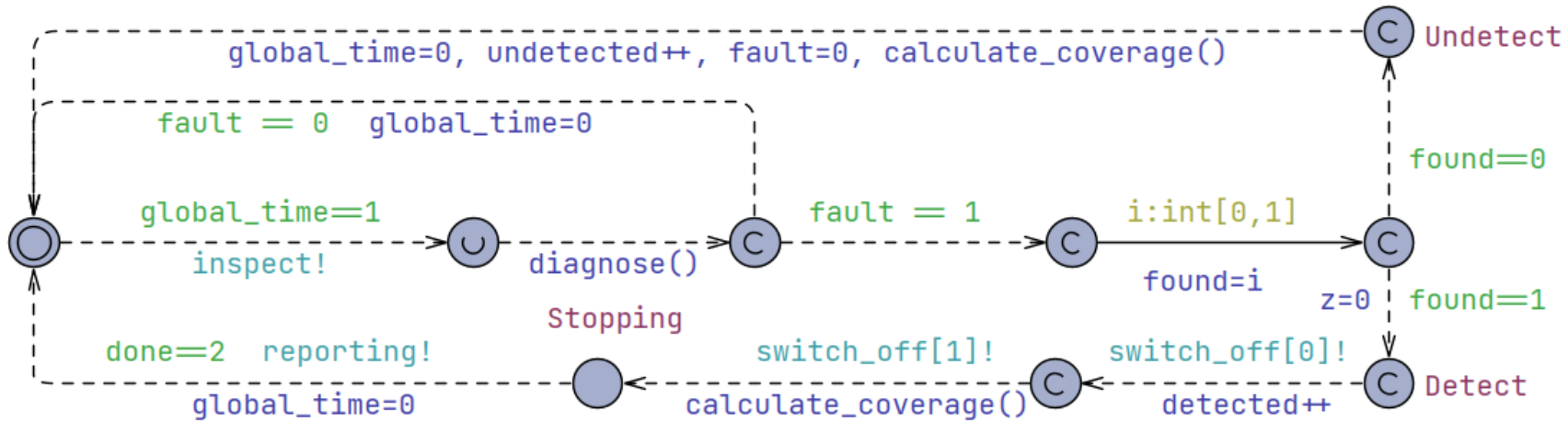}
        \caption{Diagnostic block model constructed in {\sc Stratego}}
        \label{fig:DB}
    \end{subfigure}
    \begin{subfigure}{\textwidth}
        \centering
        \includegraphics[width=0.60\linewidth]{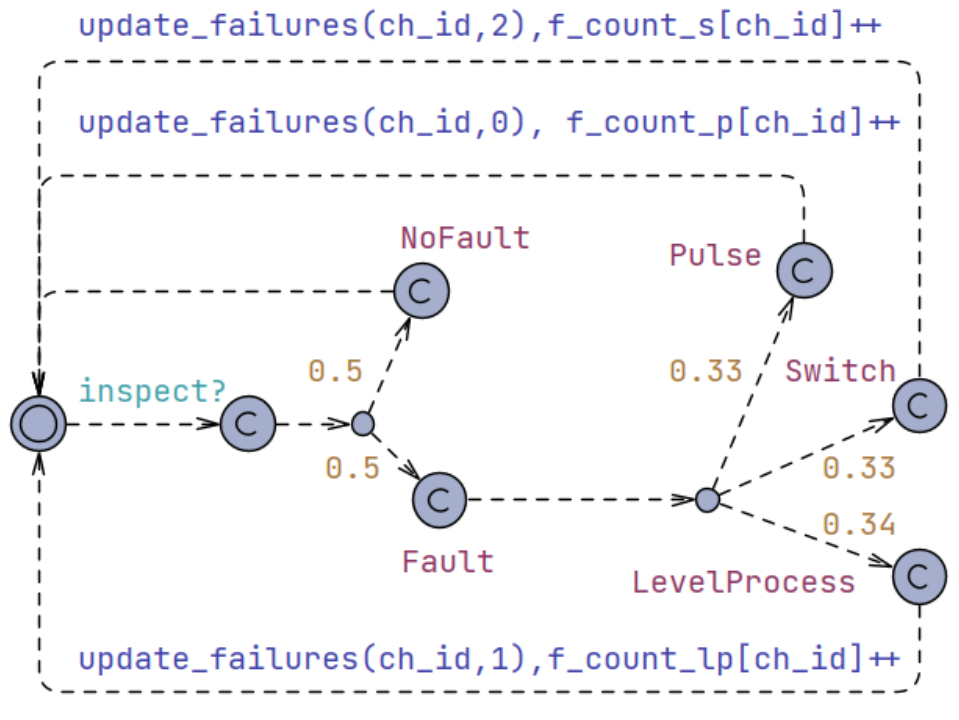}
        \caption{Fault model constructed in {\sc Stratego}}
        \label{fig:faults}
    \end{subfigure}
    \caption{Overview of the main blocks of the BDM modelled in {\sc Stratego}}
    \label{fig:overview_all}
\end{figure*}

As described in section~\ref{meth}, the formal translation of the overall BDM behaviour forms a network of TAs, where each block in Figure~\ref{fig:overview} is represented by a separate TA. In addition to BDM blocks, extra TAs are introduced to model the user interface and fault generation functionality. Since most TAs represent straightforward functionalities described in section~\ref{case}, we present only the complex and interesting ones in Figure~\ref{fig:overview_all}. To view the full model, please refer to \cite{bdmmodels}.

In Figure~\ref{fig:overview_all}, dotted lines represent controllable transitions, which are triggered by the {\sc Stratego}, while dashed lines represent uncontrollable transitions, which occur automatically based on the system's internal conditions or timing constraints. Double circles represent initial locations, single circles represent regular locations, and locations marked with ``C'' are committed, while ``U'' denotes urgent locations. Both committed and urgent locations are used for atomic transitions (no delay occurs), with committed locations having priority when enabled simultaneously.

The first transition of the \textit{DiagnosticBlock} can occur only once per minute, triggered when the guard \uppGuard{global_time == 1} is satisfied, representing that diagnostics are performed every minute (see Figure~\ref{fig:DB}).
At this point, \textit{DiagnosticBlock} communicates with the one of the two \textit{Fault} automata (depicted in Figure \ref{fig:faults}) via the channel \uppChan{inspect}. 
The \textit{Fault} template is responsible for generating faults in both STO Ch\_A and STO Ch\_B channels. 
Committed and urgent locations ensure correct information flow. As the \textit{DiagnosticBlock} is at an urgent location and a \textit{Fault} is at a committed location, the \textit{DiagnosticBlock} has to wait for all committed transitions in the \textit{Fault} to complete. The urgent location also ensures control returns to the \textit{DiagnosticBlock}, not some regular location of any other block.
The system is free to select any channel randomly for fault generation in any of its blocks.
We modelled an equal probability (dashed line transitions from a small circle before the \uppLoc{Fault} and \uppLoc{NoFault} locations) for the system to encounter a fault or remain fault-free in each diagnostic inspection cycle. If the system opts not to introduce a fault during a cycle, it transitions back to the initial location via the \uppLoc{NoFault} location. Conversely, if a fault is introduced, the system moves to the \uppLoc{Fault} location.
From this location, the system can introduce a fault in the pulse generator, switch, or SLP blocks (through \uppLoc{Pulse}, \uppLoc{Switch}, or \uppLoc{LevelProcess} locations)—with equal probability.
In the subsequent transition, the fault count for the respective block is incremented, and the overall faults record is updated through the \uppUpdate{update_failures()} function.

Now, the second transition in the \textit{DiagnosticBlock} gets enabled, triggering the function \uppUpdate{diagnose()} to check for new faults in the channels. If no faults are found in the fault record, the guard \uppGuard{fault==0} is validated, and the system returns to the initial state after resetting \uppUpdate{global_time} variable. Conversely, if faults are found in the fault record, the guard \uppGuard{fault==1} is satisfied. During the subsequent transition, the statement \uppSelect{i:int[0,1]} assigns a temporary variable \uppVar{i} a random value of 0 or 1, which is then used to update the variable \uppUpdate{found=i}. This approach represents the randomness in ability of DB to either detect or miss faults. During SFF optimization, {\sc Stratego} considers the whole system as an SHG, carefully selecting random values that enhance SFF during strategy learning. 

The guard \uppGuard{found == 0} indicates that the DB missed the fault, causing the system to transition to the \uppLoc{undetect} location.
In the subsequent transition, \uppUpdate{global_time} and \uppUpdate{fault} are reset, the \uppUpdate{undetected} counter is incremented, and the function \uppUpdate{calculate_coverage()} updates the diagnostic coverage~(DC).
A detailed description of DC is provided in Section~\ref{DO}.

On the other hand, when \uppGuard{found==1} gets valid, the system reaches the \uppLoc{detect} location, indicating fault detection. In the next transition, the counter \uppUpdate{detected} increments, and DB synchronizes with \textit{Switch(A)} via the \uppChan{switch_off[0]} channel. Similarly, in the following transition, \uppUpdate{calculate_coverage()} updates the current SFF, and DB synchronizes with \textit{Switch(B)} via \uppChan{switch_off[1]}.

After a 5\,ms startup delay, the switches close and send power output signals to \textit{PowerProcessing(A)} and \textit{PowerProcessing(B)}. Then, after another 5\,ms delay, both power processing blocks send a power-off signal to the motor. Once both signals are received, the motor shuts down.

The next transition from \uppLoc{Stopping} occurs only when \uppGuard{done==2} is satisfied, with \uppVar{done} set by the motor upon shutdown.
From \uppLoc{Stopping}, the DB sends \uppChan{reporting!} signal to the \textit{UserInterface} and \textit{Initializer} TAs, reporting the fault. 
The operator cuts power to the STO channels through the \textit{Initializer}, analyzes and resolves the fault, and then reactivates the STO channels through the \textit{Initializer} to resume normal operation.

\subsection{Safety Formal Verification Evaluation}\label{VR}

We incorporate startup delay requirements for BDM blocks as timing constraints on locations during formal modelling. Now, we identify key functional and safety requirements for BDM's operation and formally verify them using CTL  queries.

\renewcommand{\arraystretch}{1.3} 
\begin{table}[t!]
\scriptsize
    \centering
    \begin{tabular}{ccc}
        \hline
         &\textbf{Query} & \textbf{Result} \\
        \hline 
        
        1&\uppProp{Switch(0).Open -> PowerProcessing(0).Inactive} & Valid \\
       
        2&\uppProp{Switch(0).Closed -> PowerProcessing(0).Active} & Valid \\
       
        3&\uppProp{PowerProcessing(0).Active and PowerProcessing(1).Active-> Motor.On} & Valid \\
        
        4&\uppProp{PowerProcessing(0).Inactive and PowerProcessing(1).Inactive -> Motor.Off} & Valid\\
       
        5&\uppProp{DiagnosticBlock.Stopping -> Motor.Off} & Valid \\

        6&\uppProp{E<> Initializer.STO and STOChannel.Active}  & Invalid\\
       
        7&\uppProp{E<> Initializer.STO and (PulseGenerator(0).Generate or PulseGenerator(1).Generate)} & Invalid \\

        8&\uppProp{E<> Initializer.STO and (Switch(0).Closed or Switch(1).Closed)} &Invalid \\
        
        9&\uppProp{E<> Initializer.STO and (PowerProcessing(0).Active or PowerProcessing(1).Active)} & Invalid\\
        
        10&\uppProp{Initializer.STO and Motor.On}  & Invalid \\
       
        11&\uppProp{A[] ((DiagnosticBlock.Stopping and m_status == 1 and z>10 and z<=28) imply Motor.Off)} & Valid \\
        
        12&\uppProp{A[] ((Motor.Off and m_status == 0 and z>10 and z<=30) imply Initializer.STO)}  &Valid \\
        \hline
    \end{tabular}
    \caption{Formal verification results}
    \label{tab:simple_table}
\end{table}

The formal verification results are shown in Table~\ref{tab:simple_table}. Queries 1–6 evaluate liveness in core functionalities, ensuring system progresses as intended. For instance, queries 1 and 2 verify that an open switch deactivates the power processing block, while a closed switch activates it. Query 3 checks that the motor turns on when both power processing blocks are active, while query 4 ensures it turns off when both are inactive. Query 5 confirms that if the DB reaches the stopping location after detecting a fault, the motor eventually shuts down.

Queries 6–12 assess the system's safety properties. Queries 6–10 verify that unintended states are never reached simultaneously, confirming the absence of undesired behaviours. None of these queries is satisfied, indicating no violations. Query 6 ensures that when STO is activated, the STO channels deactivate and stop sending signals to the pulse generator. Queries 7–10 confirm that in STO mode, switches remain open, power processing blocks stay inactive, and the motor keeps turned off.

Query 11 verifies the timing budget requirement, ensuring the motor turns off within 28\,ms after the DB detects a fault and initiates the STO call. Similarly, Query 12 checks that STO activation occurs within 30\,ms of fault detection and the STO call. 

\subsection{SFF Optimization Evaluation}\label{DO}
Using the formally verified correct model as a base, we now focus on optimizing SFF and evaluate BD's ability to detect failures, as given in Equation~\ref{EQSFF}:
\begin{align}
        SFF=& \frac{\lambda_S + \lambda_{DD}}{\lambda_S + \lambda_{DD}+ \lambda_{DU}}\label{EQSFF}
\end{align}
\noindent where $\lambda_S$ represents the number of detected safe failures, $\lambda_{DD}$ denotes the number of detected dangerous failures, and $\lambda_{DU}$ shows the number of undetected dangerous failures.
 In this work, we disregard safe failures (i.e., \( \lambda_S = 0 \)) in the system as they do not compromise safety but cause spurious activation of the STO. For example, a faulty voltage or temperature sensor may incorrectly detect an over voltage or over temperature condition, causing the motor to stop.

Recall that the DB randomly determines whether to detect or miss faults. Without an optimization mechanism, this randomness can compromise SFF, leading to unpredictable system reliability. In order to achieve 90\,\% SFF, the DB must detect at least 90\,\% of dangerous failures. 

In order to provide {\sc Uppaal Stratego} with an optimizing criteria, we formulated a fitness function (F, see Equation~\ref{lab:fitness}). During the optimization process, this formula enables {\sc Uppaal Stratego} to compute F over the period \( t_0 \) to \( t_n \), given that the \text{SFF} and \( \text{SFF}_{\text{target}} \) are known throughout this duration.
\begin{equation}
F(t_0, t_n) = \int_{t_0}^{t_n} abs\left(\mathit {SFF(t)-SFF_{target}} \right) \, dt
\label{lab:fitness}
\end{equation}
\noindent In Equation~\ref{lab:fitness}, SFF represents the current value at time $t$, updated each time the DB detects or misses a fault, while SFF$_{target}$  is the target value (90 in this case). The objective is to minimize the absolute difference between them. The fitness function applies higher penalties as the gap between them increases and vice versa. In other words, when DB detects more faults, SFF improves (see Equation~\ref{EQSFF}), reducing the gap.
We use the following query to synthesize a strategy:
\begin{align}\label{StrategyQuery}
    \uppProp{strategy S = minE(F)[<=10000]\{\}->\{SFF\}: <> t==10000}
\end{align}
which finds a strategy that constantly observes current SFF and tries to minimize F over a period of 10,000 minutes. Since fault generation and their detection mechanism occur every minute, a period of 10,000 minutes is considered sufficient to generate adequate number of faults and evaluate how effectively the system detects them. Once the strategy is synthesized, Equations~\ref{StrategyQuery1}-\ref{StrategyQuery2} compute number of detected and undetected faults under strategy S.
\begin{align}
    \uppProp{E[time<=10000; 1] (max:detected) under S}\label{StrategyQuery1}\\
    \uppProp{E[time<=10000; 1] (max:undetected) under S}\label{StrategyQuery2}
\end{align}

\begin{figure}[!t]
    \centering
    \includegraphics[width=1.0\linewidth]{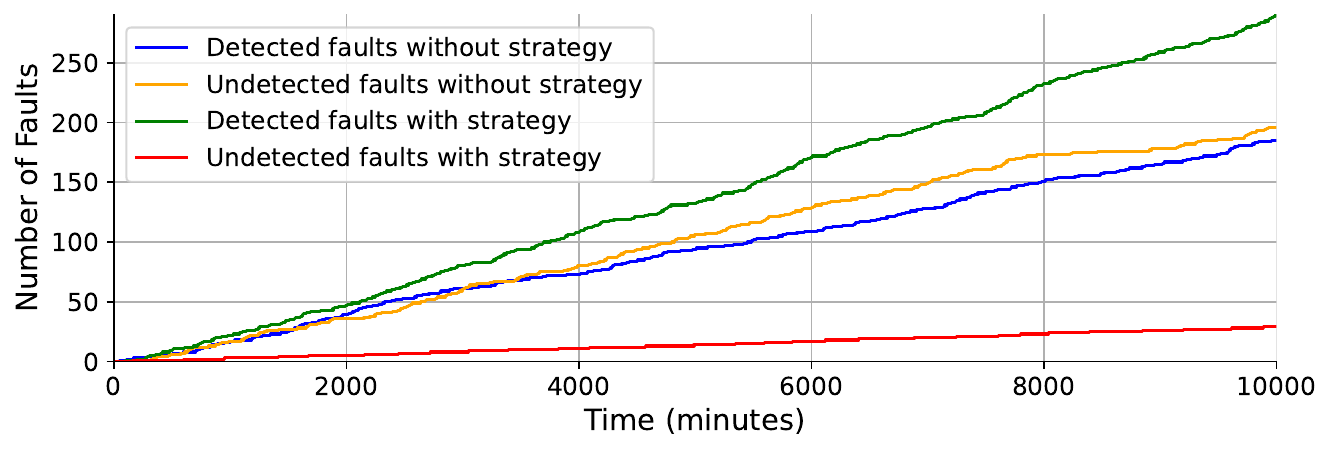}
    \caption{Fault detection with and without a strategy}
    \label{fig:faultdetection}
\end{figure}

Figure~\ref{fig:faultdetection} presents experimental results for fault detection performed with and without a strategy. Without a strategy, only 185 faults are detected, while 196 are missed, yielding unacceptable SFF of 48.5\,\% ($\frac{185}{185+196}$). In contrast, with the strategy, DB detects 290 faults, missing only 29, achieving an SFF of 90.9\,\% ($\frac{290}{290+29}$) and meeting the threshold ($\geq$90\,\%). The results highlight how Q-learning in {\sc Stratego} helps DB to enhance its fault detection performance and improve overall system reliability.

\section{Related Work}\label{rel}

Formal verification provides formal guarantees that a system functions correctly. Paul~et.~al.~\cite{FV1} applied it to aerospace applications for addressing their safety-critical issues through runtime evaluations. Similarly, Guha~et.~al.~\cite{FV3} applied it to an airbag system, demonstrating its effective role in identifying and mitigating safety-critical faults. Other studies~\cite{FV2, 10501932, Hasrat2018FormalSA} further validate its impact on flight-critical software, networks, and distributed protocols.

 Studies~\cite{WANG2020115036, LISSA2021100043} investigated how RL can optimize varying energy demands in building controls, smart energy management for heating and hot water use, and reduce CO$_2$ emissions in smart homes. {\sc Uppaal Stratego} (with an embedded RL facility) has been used in various applications for strategy synthesis, performance analysis, and optimal control. These applications include optimal scheduling for batteries~\cite{kristjansen2023dual}, swarm robotics~\cite{bogh2022distributed}, storm water detention ponds~\cite{10591152}, traffic light control~\cite{stratego_app2}, optimal railway system control~\cite{karra2019safe}, adaptive cruise control~\cite{stratego_app5}, and domestic heating control~\cite{HASRAT2023102987,online_compositional,10.1007/978-3-031-43681-9_7}.

The works mentioned above demonstrate how formal verification and RL optimization techniques are individually useful in addressing real-world problems. However, in our work, we demonstrate how formal verification and RL can be effective in a combined framework setting.

\section{Conclusion and Discussion}\label{dis}

In this paper, we present a detailed approach to addressing the safety and diagnostic challenges of industrial drive systems through a combination of formal modelling, verification, and optimization techniques. We demonstrate our approach using a real-world industrial case provided by Danfoss. A comprehensive formal model of the system was developed, capturing its functional behaviour, safety functions, and timing constraints, using  {\sc Uppaal}. Formal verification was conducted to ensure that the system does not exhibit any undesirable behaviour and satisfies key safety properties. Additionally, a reinforcement learning technique from {\sc Uppaal Stratego} was applied to the diagnostic block to achieve the target safe failure fraction and diagnostic coverage required by the design specifications of iSM3 products. The optimization approach adopted in this work not only seamlessly integrates with formal verification but also achieves a safe failure fraction of over 90.9\,\%.

To perform SFF optimization on a correct model, we first formally verify it. However, the model's continuous nature and large state space make verification in Section~\ref{VR} challenging. To address this, we introduce a deadlock condition by applying a guard on the global clock, halting the BDM system at a specific time point, thereby limiting the state space and making formal verification possible. This condition is enforced on the STO output transition, cutting off DC power to the pulse generator when the global time exceeds 100 minutes. BDM completes one cycle per minute, making the 100-minute limit sufficient to capture system behaviour and observe fault generation and detection.
For SFF optimization, we modify the formally verified model by removing the deadlock condition, marking transitions as uncontrollable if they are not directly controlled by {\sc Stratego}, and assigning random weights to locations without invariants or untriggered transitions, as required by {\sc Stratego} for the simulations to progress during strategy synthesis. In our current experiments, we synthesize a strategy for 10000 minutes, which provides a sufficient learning period for {\sc Stratego} to achieve a targeted SFF threshold of 90\,\%. Although we introduced a deadlock condition for verification, the SFF optimization problem is handled without it.

For future work, we aim to adopt some suitable techniques to address the state space explosion problem. We also aim to extend our approach to optimize the energy efficiency and operating costs of the BDM. Energy efficiency can be improved by minimizing energy waste during low-load or idle conditions while maintaining safety criteria. Operating costs can be reduced through predictive maintenance, preventing faults before they become serious and lowering life-cycle expenses.

\bibliographystyle{splncs04} 
\bibliography{References}
\end{document}